\documentclass[ALICE,manyauthors]{cernphprep}
\newif\ifcernpp
\cernpptrue

\usepackage{mathrsfs} 
\usepackage[modulo]{lineno}
\usepackage{graphicx}
\usepackage[loose]{units}
\usepackage{upgreek}
\usepackage{amssymb}
\usepackage{color}
\usepackage[T1]{fontenc}

\ifcernpp
\usepackage{cite}
\usepackage[comma,square,numbers,sort&compress]{natbib}
\usepackage{hyperref}
\else
\smartqed  
\usepackage{mathptmx}      
\usepackage{flushend}
\usepackage[numbers,sort&compress]{natbib}
\usepackage[colorlinks,draft,citecolor=blue,urlcolor=blue,linkcolor=blue]{hyperref}
\journalname{Eur. Phys. J. C}
\fi

\ifcernpp
\else
\fi

\newcommand{\tn}[1]{\textnormal{#1}}

\newcommand{\TeV}[0]{\tn{ TeV}}

\newcommand{\pPb}{\textnormal{p--Pb}}
\newcommand{\dAu}{\textnormal{d--Au}}
\newcommand{\pA}{\textnormal{p--}A}
\newcommand{\RpPb}{\ensuremath{R_\mathrm{pPb}}}
\newcommand{\QpPb}{\ensuremath{Q_\mathrm{pPb}}}
\newcommand{\PbPb}{\textnormal{Pb--Pb}}

\newcommand{\pp}{\ensuremath{\mbox{p}\mbox{p}}}
\newcommand{\snn}{\ensuremath{\sqrt{s_\mathrm{NN}}}}
\newcommand{\kt}{\ensuremath{k_\mathrm{T}}}
\newcommand{\pt}{\ensuremath{p_\mathrm{T}}}\newcommand{\pT}{\pt}
\newcommand{\ptch}{\ensuremath{p_\mathrm{T,\,ch\;jet}}}

\newcommand{\npart}{\ensuremath{N_\mathrm{part}}}

\newcommand{\ncoll}{\ensuremath{N_\mathrm{coll}}}

\newcommand{\deltaptch}{\ensuremath{\delta p_\mathrm{T,\,ch}}}

\begin{document}
%
%

\newcommand{\CKBNOTE}[1]{{\bf CKB:  #1}} 

\newcommand{\RHNOTE}[1]{{\bf RH:  #1}} 

\ifcernpp
\begin{titlepage}
\PHyear{2016}
\PHnumber{052}                 
\PHdate{29 February}              
\else
%
\fi
%
%

\title{Centrality dependence of charged jet production\\in p--Pb collisions at $\sqrt{s_\mathrm{NN}}$ = 5.02 TeV}

\ifcernpp
\ShortTitle{Centrality-dependent charged jets in p--Pb}   
%
\Collaboration{ALICE Collaboration%
         \thanks{See Appendix~\ref{app:collab} for the list of collaboration
                      members}}
\ShortAuthor{ALICE Collaboration}      
\else
\author{ALICE Collaboration}
\maketitle
\fi

\begin{abstract}

Measurements of charged jet production as a function of centrality are presented for $\pPb$ collisions recorded at $\snn = 5.02$ TeV with the ALICE detector. 
Centrality classes are determined via the energy deposit in neutron calorimeters at zero degree, close to the beam direction, to minimise dynamical biases of the selection. 
The corresponding number of
participants or binary nucleon--nucleon collisions is
determined based on the particle production in the Pb-going rapidity
region.
Jets have been reconstructed in the central rapidity region from
charged particles with the anti-$\kt$ algorithm for resolution
parameters $R = 0.2$ and $R = 0.4$ in the transverse momentum range 20 to 120 GeV/$c$.
The reconstructed jet momentum and yields have been corrected for detector effects and underlying-event background. 
In the five centrality bins considered, the charged jet production in $\pPb$ collisions is consistent with the production expected from binary scaling from pp collisions. 
The ratio of jet yields reconstructed with the two different
resolution parameters is also independent of the centrality
selection, demonstrating the absence of major modifications of the
radial jet structure in the reported centrality classes.
\end{abstract}

\ifcernpp
\end{titlepage}
\setcounter{page}{2}
\else
\fi
%

\section{Introduction}


The measurement of benchmark processes in proton--nucleus collisions plays a
crucial role for the interpretation of
nucleus--nucleus collision data, where one expects to create a system with high temperature in which the elementary constituents of hadronic matter, quarks and gluons, are deconfined for a short time: the quark-gluon plasma (QGP) \cite{Shuryak:1980tp}.
Proton--lead collisions are important to investigate cold nuclear initial and final state effects, in particular to disentangle them from effects of the hot medium created in the final state of $\PbPb$ collisions \cite{Salgado:2011wc}. 

The study of hard parton scatterings and their subsequent fragmentation via reconstructed jets plays a crucial role in the characterisation of the hot and dense medium produced in $\PbPb$ collisions while jet measurements in $\pPb$ and $\pp$ collisions provide allow to constrain the impact of cold nuclear matter effects in heavy-ion collisions.
In the initial state, the nuclear parton distribution functions can be modified with respect to the quark and gluon distributions in free nucleons,  e.g.\ via shadowing effects and gluon saturation \cite{Salgado:2011wc,McLerran:2001sr}. 
In addition, jet production  may be influenced, already in $\pPb$ collisions, by multiple scattering of partons and
hadronic re-interaction in the initial and final state
\cite{Krzywicki:1979gv,Accardi:2007in}.

In the absence of any modification in the initial state, the partonic
scattering rate in nuclear collisions compared to $\pp$ collisions is expected to
increase linearly with the average number of binary nucleon--nucleon collisions $\left< \ncoll
\right>$. 
This motivates the definition of the nuclear modification factor $\RpPb$,
 as the ratio of particle or jet transverse momentum ($\pt$) spectra in nuclear collisions to those in
$\pp$ collisions scaled by $\left< \ncoll
\right>$. 

In heavy-ion collisions at the LHC, binary ($\ncoll$) scaling is found to hold for
probes that do not interact strongly, i.e.\ isolated prompt
photons \cite{Chatrchyan:2012vq} and electroweak bosons
\cite{Chatrchyan:2011ua,Aad:2012ew}. 
On the contrary, the yields of hadrons and jets in central $\PbPb$
collisions are strongly modified compared to the scaling
assumptions. 
For hadrons, the yield is suppressed
by up to a factor of seven at $\pt \approx 6$~GeV$/c$, approaching a factor of two at
high $\pt$ ($\gtrsim$ 30 GeV/$c$) \cite{Aamodt:2010jd,Aamodt:2011vg,CMS:2012aa}.
A similar suppression is observed for jets \cite{Aad:2010bu,Chatrchyan:2012nia,Aad:2012vca,Abelev:2013kqa,Aad:2014bxa}.
This observation, known as jet quenching, is attributed to
the formation of a QGP in the collision, where
the hard scattered partons radiate gluons due to strong interaction with
the medium, as first predicted in \cite{Gyulassy:1990ye,Baier:1994bd}.

In minimum bias $\pPb$ collisions at $\snn = 5.02$~TeV the production
of unidentified charged particles
\cite{ALICE:2012mj,Abelev:2014dsa,Khachatryan:2015xaa,ATLAS-CONF-2014-029}
and jets \cite{ATLAS:2014cpa,Chatrchyan:2014hqa,Adam:2015hoa} is
consistent with the absence of a strong final state suppression.
However, multiplicity dependent studies in $\pPb$ collisions on the production of low-$\pt$ identified particles and
long range correlations  \cite{CMS:2012qk,Abelev:2012ola,Aad:2013fja,Abelev:2013wsa} show similar features as measured
in $\PbPb$ collisions, where they are attributed to the collective
behaviour following the creation of a QGP.  
These features in $\pPb$  collisions become more pronounced for higher multiplicity events, which in $\PbPb$ are commonly associated with more central collisions or
higher initial energy density. 

The measurement of jets, compared to single charged hadrons, tests the
parton fragmentation beyond the leading particle with the inclusion of
large-angle and low-$\pt$ fragments. 
Thus jets are potentially sensitive to centrality-dependent
modifications of low-$\pt$ fragments. 

This work extends the analysis of the charged jet production in minimum bias $\pPb$ collisions recorded with the ALICE detector at $\snn =
5.02$ TeV to a centrality-differential study for jet resolution parameters $R = 0.2$ and 0.4  in the $\pt$ range from 20 to
120 GeV/$c$ \cite{Adam:2015hoa}.
Section~\ref{sec:analysis} describes the event and  track selection, the centrality determination, as well as the jet reconstruction, the corrections 
for uncorrelated background contributing to the jet momentum \cite{Cacciari:2011tm,Abelev:2013kqa,Abelev:2012ej} and the corrections for detector effects. 
The impact of different centrality selections on the nuclear modification factor has been studied in detail in \cite{Adam:2014qja}. 
We estimate the centrality using zero-degree neutral energy and the charged particle multiplicity
measured by scintillator array detectors at rapidities along the
direction of the Pb beam to determine $\ncoll$.   
The correction procedures specific to the centrality-dependent jet measurement are discussed in detail.
Section~\ref{sec:obs} introduces the three main observables: the centrality-dependent jet production cross section, the nuclear modification factor, and ratio of jet cross sections for two different resolution parameters. 
Systematic uncertainties are discussed in Sec.~\ref{sec:syserr} and results are presented in Sec.~\ref{sec:results}. 

\section{Data analysis}
\label{sec:analysis}

\subsection{Event selection}

The data used for this analysis were collected with the ALICE detector
\cite{Abelev:2014ffa} during the $\pPb$ run of the LHC at
$\sqrt{s_\mathrm{NN}} = 5.02 \TeV$ at the beginning of
2013. The ALICE experimental setup and its performance during the LHC
Run 1 are described in
detail in \cite{Aamodt:2008zz,Abelev:2014ffa}.

For the analysis presented in this paper, the main detectors used for event and centrality selection are two
scintillator detectors (V0A and V0C), covering the
pseudo-rapidity range of $2.8 < \eta_\mathrm{lab}< 5.1$ and  $-3.7 <
\eta_\mathrm{lab} < -1.7$, respectively \cite{Aamodt:2010pb}, and the Zero
Degree Calorimeters (ZDCs), composed of two sets of neutron (ZNA and ZNC) and
proton calorimeters (ZPA and ZPC) located at a distance $\pm112.5$~m from the
interaction point. 
Here and in the following $\eta_\mathrm{lab}$ denotes the
pseudo-rapidity in the ALICE laboratory frame.

The minimum bias trigger used in $\pPb$ collisions requires signal coincidence in the V0A and V0C scintillators. 
In addition, offline selections on timing and vertex-quality are used to remove events with multiple interactions within the same bunch crossing and (pile-up) and background events, such as beam-gas interactions. 
The event sample used for the analysis presented in this manuscript
was collected exclusively in the beam configuration where the proton
travels towards negative $\eta_\mathrm{lab}$ (from V0A to V0C).   
The nucleon--nucleon center-of-mass system moves in the direction of
the proton beam corresponding to a rapidity
of $y_\mathrm{NN} = -0.465$.  

A van der Meer scan was performed to measure the visible cross section for
the trigger and beam configuration used in this analysis: $\sigma_\mathrm{V0} = 2.09 \pm
0.07$~b  \cite{Abelev:2014epa}. 
Studies with Monte Carlo simulations show that the sample collected in the configuration explained above consists mainly of non-single
diffractive (NSD) interactions and a negligible contribution from
single diffractive and electromagnetic interactions (see
\cite{ALICE:2012xs} for details). 
The trigger is not fully efficient for NSD events and the inefficiency
is observed mainly for events without a reconstructed vertex,
i.e.\ with no particles produced at central rapidity. 
Given the fraction of events without a reconstructed vertex in the data the corresponding inefficiency
for NSD events is estimated to ($2.2 \pm 3.1$)\%.
This inefficiency is expected to mainly affect the most peripheral centrality class.
Following the prescriptions of \cite{Adam:2014qja}, centrality classes are defined as
percentiles of the visible cross section and are not corrected for
trigger efficiency.

The further analysis requires a reconstructed vertex, in addition to the minimum bias trigger selection.
The fraction of events with a reconstructed vertex is  98.3\% for minimum bias events and depends on the centrality class.
In the analysis events with a reconstructed vertex $|z| > 10~\mathrm{cm}$ along the beam axis are
rejected. 
In total, about $96\cdot10^6$ events, corresponding to an integrated luminosity of 46 $\mu$b${}^{-1}$, are used for the analysis and classified into five centrality classes 

\subsection{Centrality determination}
\label{sec:cent}
Centrality classes can be defined by dividing the 
multiplicity distribution measured  in a certain pseudo-rapidity interval into
fractions of the cross section, with the highest multiplicities
corresponding to the most central collisions (smallest impact parameter $b$).
The corresponding number of  participants, as well as $\ncoll$ and
$b$, can be
estimated with a Glauber model \cite{Miller:2007ri}, e.g.\ by fitting the
measured multiplicity distribution with the $\npart$ distribution from
the model, convoluted with a Negative Binomial Distribution (NBD). 
Details on this procedure for $\PbPb$ and $\pPb$ collisions in ALICE are found in
\cite{Abelev:2013qoq} and \cite{Adam:2014qja}, respectively.

In $\pA$ collisions centrality selection is susceptible to a variety of biases.
In general, relative fluctuations of $\npart$ and of event multiplicity are
large, due to their small numerical value, in $\pPb$ collisions  \cite{Adam:2014qja} $\left< \npart \right>  = \left< \ncoll \right> + 1 = 7.9 \pm 0.6$ and 
$\frac{\mathrm{d}N_\mathrm{ch}}{\mathrm{d}\eta} = 16.81 \pm 0.71$, respectively.
Using either of these quantities to define centrality, in the Glauber model or the in
experimental method, already introduces a bias compared to a purely geometrical selection based on the impact
parameter $b$. 

In addition, a kinematic bias exists for events containing high-$\pT$
particles, originating from parton fragmentation as discussed above.
The contribution of these jet fragments to the overall multiplicity
rises with the jet energy and thus can introduce a trivial correlation
between the multiplicity and presence of a high-$\pt$ particle, and a
selection on multiplicity will bias the jet population.
High multiplicity events are more likely created in
collisions with multiple-parton interactions, which can lead to a nuclear
modification factor larger than unity.
On the contrary, the selection of low multiplicity (peripheral) events
can pose an effective veto on hard processes, which would lead to a nuclear
modification factor smaller than unity.
As shown in \cite{Adam:2014qja} the observed suppression and
enhancement for charged particles
in bins of multiplicity with respect to the binary scaling assumption can be explained by this selection bias alone.
The bias can be fully reproduced by an independent superposition of simulated $\pp$ events and the farther the centrality
estimator is separated in rapidity from the measurement region at
mid-rapidity, the smaller the bias. 
We do not repeat the analysis for the centrality estimators with known biases here.

In this work, centrality classification is based solely on the zero-degree energy measured in the lead-going neutron detector ZNA, since it is
expected to have only a small dynamical selection bias. 
However, the ZNA signal cannot be related directly to the produced multiplicity for the $\ncoll$ determination via NBD. 
As discussed in detail in  \cite{Adam:2014qja} an alternative hybrid approach is used to connect the centrality
selection based on the ZNA signal to another $\ncoll$ determination via the charged particle
multiplicity in the lead-going direction measured with the V0A ($\left< \ncoll \right>_c^\mathrm{Pb-side}$). 
This approach assumes that the V0 signal is proportional to the number of
wounded lead (target) nucleons ($N_\mathrm{part}^\mathrm{target} =
\npart - 1 = \ncoll$).
The average number of collisions for a given centrality, selected with
the ZNA, is then given by scaling the minimum bias value $\left< \ncoll
\right>_\mathrm{MB} = 6.9$ with the ratio of the average raw signal
$S$ of the innermost ring of the V0A:
\begin{equation}  
\left< N_\mathrm{coll}^\mathrm{Pb-side} \right>_c = \left< \ncoll
\right>_\mathrm{MB} \cdot \frac{\left<S \right>_c}{\left<S\right>_\mathrm{MB}}. 
\end{equation}
The values of $\ncoll$ obtained with this method are shown in Tab.~\ref{tab:cents_moments} for different ZNA centrality classes \cite{Adam:2014qja}. 

\begin{table}
\centering
\ifcernpp
\footnotesize
\else
\fi
\begin{tabular}{l r@{$\pm$}lccc  }
\hline
ZNA centrality class & \multicolumn{2}{c}{{$\left<N_\mathrm{coll}^\mathrm{Pb-side}\right>$}} &  $\rho$ & $\sigma(\rho)$ & $\sigma(\deltaptch) (R = 0.4)$ \\
(\%) of visible cross section &  \multicolumn{2}{c}{} & (GeV/$c$) &  (GeV/$c$) & (GeV/$c$) \\
\hline
0-20   & 12.1 & 1.0  &1.60 & 1.17 & 1.43\\
20-40 & 9.6 & 0.8 & 1.27 & 1.04 & 1.30 \\
40-60 & 6.7 & 0.5 & 0.88 & 0.84 & 1.11\\
60-80 & 4.0 & 0.3 & 0.70 & 0.52 &0.90\\
80-100  & 2.1 & 0.3& 0.26 & 0.37 & 0.71\\
Minimum bias (0-100) & 6.9 & 0.6 & 0.98 & 1.02 & 0.91 \\
\hline
\end{tabular} 
\caption{\label{tab:cents_moments}
 Average $\ncoll$ values  for centrality classes selected with the ZNA determined with the hybrid approach ($N_\mathrm{coll}^\mathrm{Pb-side}$)  \cite{Adam:2014qja},
as well as moments of the background density and background fluctuation distributions shown in Fig.~\ref{fig:background} (negligible statistical uncertainty).}
\end{table}

\subsection{Jet reconstruction and event-by-event corrections}
\label{sec:jetreco}

The reported measurements are performed using \emph{charged jets}, clustered starting from charged particles only, as described in  \cite{Abelev:2013kqa,ALICE:2014dla,Adam:2015hoa} for different collision systems.
Charged particles are reconstructed using information from the Inner Tracking
System (ITS) \cite{Aamodt:2010aa} and the Time Projection Chamber
(TPC) which cover the full azimuth and $|\eta_\mathrm{lab}| < 0.9$
 for tracks reconstructed with full length in the
TPC \cite{Alme:2010ke}. 

The azimuthal distribution of high-quality tracks with reconstructed track points in the Silicon Pixel Detector (SPD), the two innermost layers of the ITS, is not completely uniform due to inefficient regions in the SPD.
This can be compensated by considering in addition tracks \textit{without} reconstructed points in the SPD. 
The additional tracks constitute approximately 4.3\% of the track sample used for analysis.
For these tracks, the primary vertex is used as an additional constraint in the track fitting to improve the momentum resolution. 
This approach yields a uniform tracking efficiency within the
acceptance, which is needed to avoid geometrical biases of the jet
reconstruction algorithm  
caused by a non-uniform density of reconstructed  tracks.
The procedure is described first and in detail in the context of jet reconstruction with ALICE in $\PbPb$ collisions \cite{Abelev:2013kqa}. 

\begin{figure}
\setlength{\unitlength}{\textwidth}
\begin{picture}(1,0.38)
\put(0,0){\includegraphics[width=0.5\unitlength]{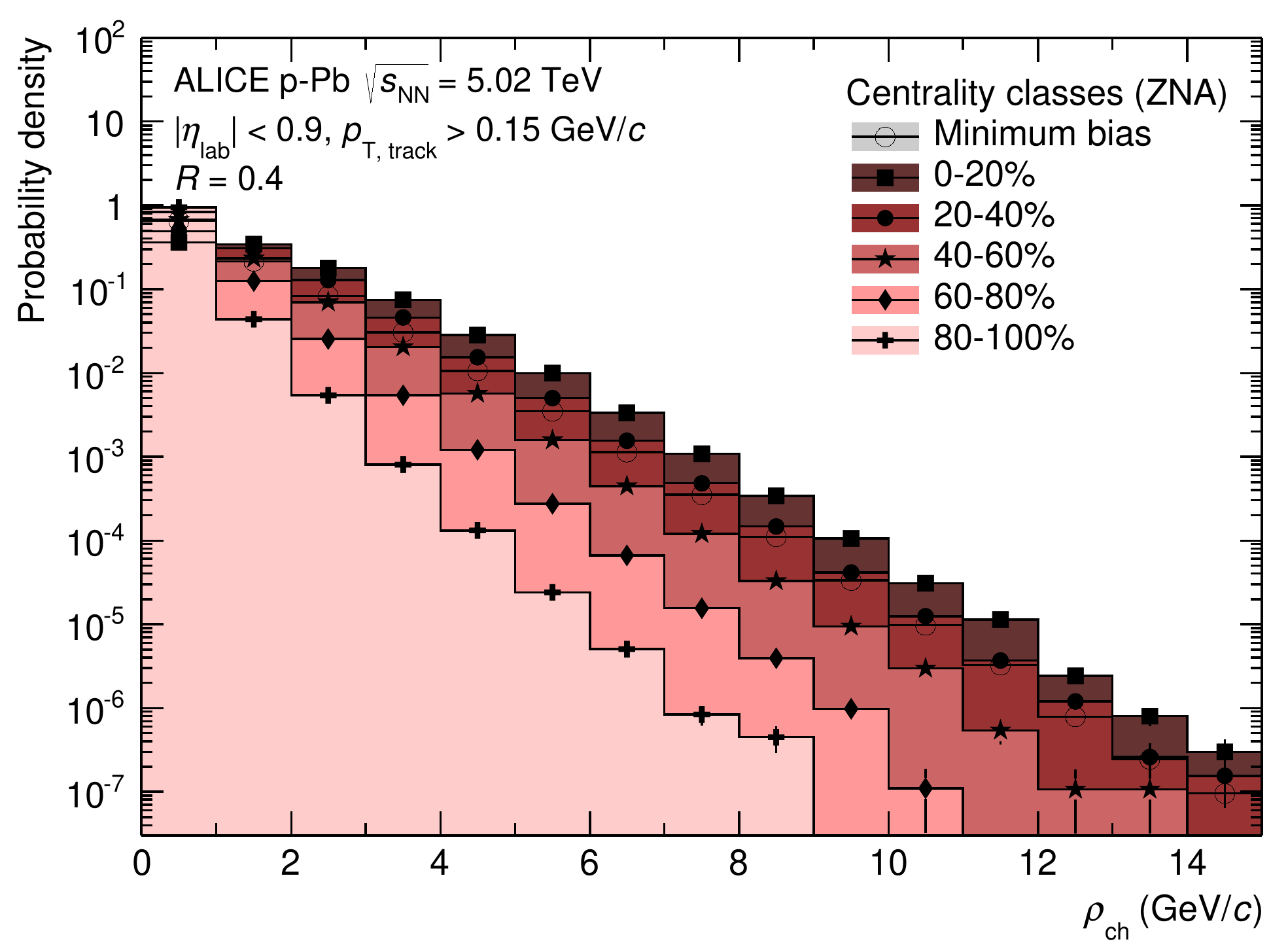}}
\put(0.5,0){\includegraphics[width=0.5\unitlength]{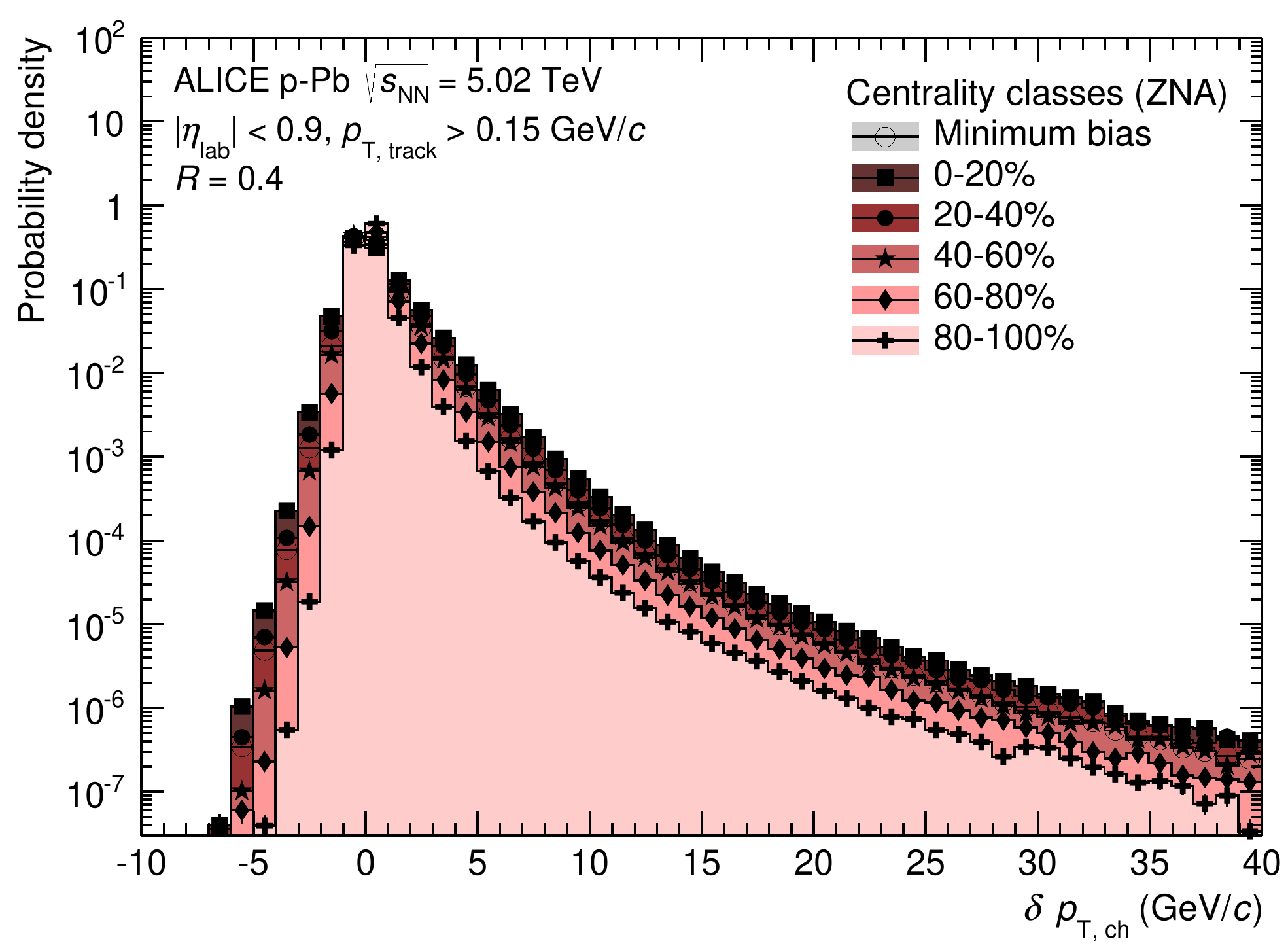}}
\end{picture}
\caption{(Color online) Left: Centrality dependence of the background
  momentum density $\rho_\mathrm{ch}$ determined with $\kt$ jets and $R = 0.4$. Right: $\deltaptch$ distributions for different
  centralities obtained with random cones and $R = 0.4$.}
\label{fig:background}
\end{figure}

The anti-$k_\mathrm{T}$ algorithm from the FastJet package
\cite{Cacciari:2008gp} is employed  to reconstruct jets from these tracks using the $\pt$ recombination scheme.
The resolution parameters used in the present analysis are $R=0.2$ and $R=0.4$.
Reconstructed jets are further corrected for contributions from the
underlying event to the jet momentum as
\begin{equation}
\label{eq:ptjet}
p_\mathrm{T,\,ch\;jet} = p_\mathrm{T,\,ch\;jet}^\mathrm{raw} -
A_\mathrm{ch\;jet} \cdot \rho_\mathrm{ch},
\end{equation} 
where $A_\mathrm{ch\;jet}$ is the area of the jet and $\rho_\mathrm{ch}$  the event-by-event background density \cite{Cacciari:2007fd}.  
The area is estimated by counting the so-called \emph{ghost particles} in the jet. 
These are defined as particles with a finite area and vanishing momentum, which are distributed  uniformly in the event and included in the jet reconstruction \cite{Cacciari:2008gn}.
Their vanishing momentum ensures that the jet momentum is not influenced when they are included, while the number of ghost particles assigned to the jet provides a direct measure of its area.
The background density $\rho_\mathrm{ch}$ is estimated via the median of
the individual momentum densities of jets reconstructed with the $\kt$ algorithm in
the event
\begin{equation}
\rho_\mathrm{ch} = \mathrm{median} \left\{ \frac{p_{\\
   \mathrm{T},\,k}}{A_k} \right\}  \cdot C,
\end{equation}
where $k$ runs over all reconstructed $\kt$ jets with
momentum $p_{\mathrm{T},\,i}$ and area $A_i$. 
Reconstructed $\kt$ jets are commonly chosen for the estimate of the background density, since they provide a more robust sampling of low momentum particles. 
$C$ is the occupancy correction factor, defined as
\begin{equation}
C = \frac{\sum_j {A_{j} }}{A_\tn{acc}},
\end{equation}
where $A_j$ is the area of each $k_\mathrm{T}$  jet with at least one real track, i.e.\ excluding ghosts, and $A_\mathrm{acc}$ is the area of the charged-particle acceptance, namely $(2 \times 0.9) \times 2\pi$.
The typical values for $C$ range from $0.72$ for most central  collisions (0-20\%)
to $0.15$ for most peripheral collisions  (80-100\%).
This procedure takes into account the more sparse environment in
$\pPb$ collisions compared to $\PbPb$ and is described in more detail
in \cite{Adam:2015hoa}.
The probability distribution for $\rho_\mathrm{ch}$ for the five 
centrality classes and minimum bias is shown in
Fig.~\ref{fig:background} (left) and the mean and width of the distributions
are given in Tab.~\ref{tab:cents_moments}. 
The event activity and thus the background density
increases for more central collisions, though on average the
background density is still two orders of magnitude smaller than in $\PbPb$
collisions where $\rho_\mathrm{ch}$ is $\approx 140$ GeV/$c$ for central collisions \cite{Abelev:2012ej}.

\subsection{Jet spectrum unfolding}
\label{sec:unfolding}

Residual background fluctuations and instrumental effects can smear the jet $\pt$. 
Their impact on the jet spectrum needs to be corrected on a statistical basis using unfolding, which is performed using the approach of
Singular-Value-Decomposition (SVD) \cite{Hocker:1995kb}.  
The response matrix employed in the unfolding is the combination of the (centrality-dependent) jet response to background fluctuations and the detector response.
The general correction techniques are discussed in detail
in the context of the minimum bias
charged jet measurement in $\pPb$  \cite{Adam:2015hoa}.

Region-to-region fluctuations of the background density
compared to the event median, contain purely statistical fluctuations of particle number and
momentum  and in addition also intra-event correlations, e.g.\ those characterised by the azimuthal anisotropy $v_2$ and higher harmonics, which induce additional variations of
the local background density. 
The impact of these fluctuations on the jet momentum is determined 
by probing the transverse momentum density in randomly distributed
cones  in $(\eta,\phi)$  and comparing it to the average background
via \cite{Abelev:2012ej}: 
\begin{equation}
\label{eq:deltapt}
\deltaptch = \sum_\mathrm{i}{p_\mathrm{T,\,i}-\rho_\mathrm{ch}\cdot A}, ~~~ A = \pi R^2
\end{equation}
where $p_\mathrm{T,\,i}$ is the transverse momentum of each track $i$ inside a cone of radius $R$, where $R$ corresponds to the resolution parameter in the jet reconstruction. 
$\rho_\mathrm{ch}$ is the background density, and $A$  the area of the cone.
The distribution of residuals, as defined by Eq.~\ref{eq:deltapt}, is
shown for different centralities in
Fig.~\ref{fig:background} (right). 
The corresponding widths are given in Tab.~\ref{tab:cents_moments}.
The background fluctuations increase for more central events, which is
expected from the general increase of statistical fluctuations ($\propto \sqrt{N}$) with the particle multiplicity. 
The  $\deltaptch$ distributions measured for $R = 0.2$ and 0.4 are used in the unfolding procedure.

In addition to the background fluctuations the unfolding procedure 
takes into account the instrumental response.
The dominating instrumental effects on the reconstructed jet spectrum are the
single-particle tracking efficiency and momentum resolution. 
These effects are encoded in a response matrix, which 
 is determined with a full detector simulation using PYTHIA6 \cite{Sjostrand:2006za} to
generate jets and GEANT3 \cite{Brun:1994aa} for the transport through the ALICE setup.
The detector response matrix links the jet momentum at the charged particle level to the
one  reconstructed from tracks after particle transport through the detector.
No correction for the missing energy of neutral jet constituents is
applied.

\section{Observables}
\label{sec:obs}
\subsection{Jet production cross sections} 

The jet production cross sections $\frac{\mathrm{d}\sigma^c}{\mathrm{d}\pT}$, for different centralities $c$, are provided as fractions of the
visible cross section $\sigma_\mathrm{V0}$. 
The fraction of the cross section is determined with the number of selected events in each centrality bin $N^c_\mathrm{ev}$ and takes into account the vertex reconstruction efficiency $\epsilon^c_\mathrm{vtx}$ determined for each centrality
 \begin{equation}
\label{eq:DefSpectraCentrality}
\frac{\mathrm{d}\sigma^c}{\mathrm{d}\pT} =
\frac{\epsilon^c_\mathrm{vtx}}{N^c_\mathrm{ev}} 
\frac{\mathrm{d}N}{\mathrm{d}\pT} \cdot \sigma_\mathrm{V0} \cdot 
\frac{N^c_\mathrm{ev}}{N^\mathrm{MB}_\mathrm{ev}} = \frac{\epsilon^c_\mathrm{vtx}}{N^\mathrm{MB}_\mathrm{ev}} \frac{\mathrm{d}N}{\mathrm{d}\pT} \cdot \sigma_\mathrm{V0},
\end{equation}
where $\epsilon^c_\mathrm{vtx}$ decreases from 99.9\% for the most central selection (0-20\%) to 95.4\% in peripheral. 

\subsection{Quantifying nuclear modification}

The nuclear modification factor compares the
$\pt$-differential per-event yield,
e.g.\ in $\pPb$ or $\PbPb$ collisions, to the differential yield in $\pp$ collisions at the 
same center-of-mass energy in order to quantify nuclear effects. 
Under the assumption that the jet or particle production at high $\pt$ scales with the number of binary collisions, the
nuclear modification factor is unity in the absence of nuclear
effects. 

In $\pPb$ collisions the jet population can be biased, depending on the
centrality selection and $\ncoll$ determination, hence the nuclear modification factor may
vary from unity even in the absence of nuclear effects as described in detail in Sec.\,\ref{sec:cent} (see also \cite{Adam:2014qja}). 
To reflect this ambiguity the centrality-differential nuclear
modification factor in $\pPb$ collisions is called $\QpPb$, instead of $\RpPb$ as in the minimum bias case. $\QpPb$ is  defined as
\begin{equation}
\label{eq:r_ppb}
\QpPb
=
\frac{
\mathrm{d^2}N^{c}_\mathrm{pPb}/\mathrm{d}\eta\mathrm{d}p_\mathrm{T}}
{\left<N_\mathrm{coll}^{c}\right>   \cdot
  \mathrm{d^2}N_\mathrm{pp}/\mathrm{d}\eta\mathrm{d}p_\mathrm{T}}.
\end{equation}
Here, $\left<N^{c}_\mathrm{coll}\right>$ is number of binary collisions for centrality $c$, shown in Tab.~\ref{tab:cents_moments}.

For the construction of $\QpPb$, we use the same $\pp$ reference as for the study of
charged jet production in minimum bias $\pPb$ collisions \cite{Adam:2015hoa}.
This reference has been determined  from the ALICE charged jet measurement at 7~TeV
\cite{ALICE:2014dla} via scaling to the $\pPb$ center-of-mass
energy and taking into account the rapidity shift of the colliding
nucleons.
The scaling behaviour of the charged jet spectra is determined based on pQCD calculations using the POWHEG framework \cite{Frixione:2007nw} and PYTHIA parton shower
 (see \cite{Adam:2015hoa} for details).
This procedure fixes the normalisation based on the measured data at 7 TeV, while the evolution of the cross section with beam energy is calculated, taking into account all dependences implemented in POWHEG and PYTHIA, e.g. the larger fraction of quark initiated jets at lower collision energy.

\subsection{Jet production cross section ratio}

The angular broadening or narrowing of the parton shower with respect to the
original parton direction can have an impact on the jet production
cross section determined with different resolution parameters. 
This can be tested via the ratio of cross sections or yields
reconstructed with different radii, e.g.\ $R = 0.2$ and $0.4$,  in a common rapidity interval, here $|\eta_\mathrm{lab}| < 0.5$:
\begin{equation}
\label{eq:xsec_ratio}
\mathscr{R}(0.2,\,0.4) = \frac{\mathrm{d}\sigma_\mathrm{pPb,\, R=0.2} / \mathrm{d}p_\mathrm{T}}{\mathrm{d}\sigma_\mathrm{pPb,\, R=0.4} / \mathrm{d}p_\mathrm{T}}.
\end{equation}

Consider for illustration the extreme scenario where all fragments are already contained within $R=0.2$. 
In this case the ratio would be unity. 
In addition, the statistical uncertainties between $R = 0.2$ and $R  = 0.4$ would be fully correlated and they would cancel completely in the ratio, when the jets are reconstructed from the same data set. 
If the jets are less collimated, the ratio decreases and the statistical uncertainties cancel only partially. 
For the analysis presented in this paper, the conditional probability varies between 25\% and 50\% for reconstructing a $R =0.2$ jet in the same $\pt$-bin as a geometrically close $R = 0.4$ jet. 
This leads to a reduction of the statistical uncertainty on the ratio of about 5-10\% compared to the case of no correlation.

The measurement and comparison of fully corrected jet cross sections for different radii provides an observable sensitive to the radial redistribution of momentum that is also theoretically well defined \cite{Dasgupta:2016bnd}.  
Other observables that test the structure of jets, such as the fractional transverse momentum distribution of jet constituents in radial and longitudinal direction or jet-hadron correlations \cite{Aamodt:2011vg,Chatrchyan:2013kwa,Chatrchyan:2014ava,Aad:2014wha,Adam:2015doa}, are potentially more sensitive to modified jet fragmentation in $\pPb$ and $\PbPb$. 
However, in these cases the specific choices of jet reconstruction parameters, particle $\pt$ thresholds and the treatment of background particles often limit the quantitative comparison between experimental observables and to theory calculations.

\section{Systematic uncertainties}
\label{sec:syserr}

The different sources of systematic uncertainties  for the three  observables presented in this paper are listed
in Tab.~\ref{tab:sys_err} for 0-20\% and 60-80\% most central collisions.

\begin{table}
\ifcernpp
\scriptsize
\else
\fi
\begin{center}
\begin{tabular}{lllllllll}
\hline
Observable & Jet cross section ($R = 0.4$) &&\phantom{xxx} &  $\QpPb$ ($R = 0.4$) &&\phantom{xxx} & $\mathscr{R}$\\
\cline{2-3}\cline{5-6}\cline{8-9}
ZNA centrality class (\%) & 0-20 & 60-80 && 0-20 & 60-80 && 0-20 & 60-80\\
\hline

Single-particle efficiency (\%) & $10.2-14.0$ & $10.0-12.7$ && $4.9-6.3$ & $4.9-6.4$ && $2.0-2.0$ & $1.8-4.7$ \\
Unfolding (\%) & 4.3 & 4.6 && 4.5 & 4.8 && 1.4 & -3.1 \\
Unfolding prior steepness (\%) & $0.9-7.0$ & $0.3-3.6$ && $1.1-7.2$ & $0.8-4.0$ && $0.7-1.4$ & $0.3-2.2$ \\
Regularisation strength (\%) & $2.8-6.4$ & $0.4-3.7$ && $2.8-7.3$ & $0.5-3.9$ && $1.8-7.0$ & $0.3-3.7$ \\
Minimum $\pt$ cut-off (\%) & $3.7-9.2$ & $0.6-2.9$ && $4.1-9.8$ & $1.7-3.8$ && $2.2-0.8$ & $0.5-1.8$ \\
Background estimate (\%) & $3.5-1.8$ & $3.8-3.0$ && $3.5-1.8$ & $3.8-3.0$ && $1.7-1.8$ & $2.6-1.2$ \\
$\deltaptch$ estimate (\%) & $0.1-0.0$ & $0.2-2.3$ && $0.1-0.0$ & $0.2-2.3$ && $0.1-0.0$ & $0.2-1.1$ \\
&&&&&&&\\
Combined uncertainty (\%) & $12.5-19.8$ & $11.6-15.2$ && $9.0-16.3$ & $8.1-11.1$ && $4.2-7.8$ & $4.4-7.5$ \\
&&&&&&&\\
Combined uncertainty ($R = 0.2$) (\%) & $10.4-19.5$ & $8.2-12.5$ && $8.6-18.0$ & $5.8-9.4$ & & - & - \\
&&&&&&&\\
$\left<N_\mathrm{coll}^\mathrm{Pb-side}\right>$ (\%) & - & - && 8.0 & 8.0 && - & - \\
Visible cross section (\%) & 3.3 & 3.3 && - & - && - & - \\
Reference scaling pp 7 TeV (\%) & - & - && 9.0 & 9.0 && - & - \\
NSD selection efficiency $\pPb$ (\%) & - & - && 3.1 & 3.1 && - & - \\ 
&&&&&&&\\
Combined scaling uncertainty (\%) & - & - && 12.4 & 12.4 && - & -\\
\hline
\end{tabular} 
\end{center}
\caption{
\label{tab:sys_err}
Summary of systematic uncertainties on the fully corrected jet
spectrum, the corresponding nuclear modification factor, and the jet
production cross section ratio in 0-20\% central and 60-80\% peripheral events for the resolution parameter $R=0.4$. 
The range of percentages provides the variation from the minimum to the maximum momentum
 in each centrality. For $R = 0.2$ only the combined uncertainty is provided for, the difference to $R = 0.4$ is mainly due to the smaller impact of the single particle efficiency for smaller radii.}
\end{table}

The dominant source of uncertainty for the $\pt$-differential jet production cross section is the uncertainty of the single-particle tracking efficiency that has a direct impact on the correction of the jet momentum in the unfolding, as discussed in Sec.\,\ref{sec:unfolding}.
In $\pPb$ collisions, the single-particle efficiency is known with a relative uncertainty of 4\%, which is equivalent to a 4\% uncertainty on the jet momentum scale.
To estimate the effect of the tracking efficiency uncertainty on the jet yield, the tracking efficiency is artificially lowered by randomly discarding the corresponding fraction  of tracks (4\%) used as input for the jet finder. 
Depending on the shape of the spectrum, the uncertainty on the single-particle efficiency (jet momentum scale) translates into an uncertainty on the jet yield ranging from 8 to 15\%.

To estimate the effect of the single-particle efficiency on the
$\pPb$ nuclear modification factor for jets,
one has to consider that the uncertainty on the efficiency is
partially correlated between the $\pp$ and $\pPb$ data set.
The correction is determined with the same description of the ALICE detector in the Monte Carlo and for similar track quality
cuts, but changes of detector conditions between run periods  reduce the degree of correlation between the data sets.
The uncorrelated uncertainty on the single-particle efficiency has been estimated to 2\% by varying the track quality cuts in data and simulations. 
Consequently, the resulting uncertainty for the nuclear modification factor is basically half the uncertainty due to the single particle efficiency in the jet spectrum (cf. Tab.\,\ref{tab:sys_err}). 
It was determined by discarding 2\% of the tracks in one of the two collision systems, as also described in \cite{Adam:2015hoa}. 

Uncertainties introduced by the unfolding procedure, e.g.\ choice of
unfolding method, prior, regularisation strength, and minimum $\pt$
cut-off, are determined by varying those methods and parameters within reasonable boundaries. 
Bayesian \cite{Blobel:2002pu,Dagostini:2003} and  $\chi^2$ \cite{Blobel1984} unfolding have been tested and compared to the default SVD unfolding to estimate the systematic uncertainty of the chosen method.
The quality of the unfolded result is evaluated by inspecting the Pearson
coefficients, where  a large (anti-)correlation between neighbouring
bins indicates that the regularisation is not optimal.

The overall uncertainty on the jet yield due to the background
subtraction is estimated by comparing various background estimates:
track-based and jet-based density estimates, as well as pseudo-rapidity-dependent corrections. 
The estimated uncertainty amounts to 3.8\% at low $\pt$ and decreases for higher reconstructed jet momenta. 

The main uncertainty related to the background fluctuation estimate is given by the choice of excluding reconstructed jets in the random cone
sampling. 
While the probability of a jet to overlap with another jet in the event
scales with $\ncoll -1$, it scales in the case of the random cone sampling with $\ncoll$.  
This can be emulated by rejecting a given fraction of cones
overlapping with signal jets, which introduces an additional dependence on the
definition of a signal jet. 
The resulting uncertainty due to the treatment of jet overlaps is of the order
of 0.1\% and can be considered negligible.

In addition, several normalisation uncertainties need to be
considered: the uncertainty on $\ncoll$ (8\% in the hybrid approach), on the
visible cross section $\sigma_\mathrm{V0}$ (3.3\%) and from the
 assumptions made to obtain the scaled pp reference from 7 to 5 TeV (9\%).

Further details on the evaluation of the centrality-independent systematic uncertainties can be
found in \cite{Adam:2015hoa}.

\section{Results}
\label{sec:results}

\begin{figure}[t]
\setlength{\unitlength}{0.75\columnwidth}
\begin{center}
\begin{picture}(1,1.25)
\put(0,0){\includegraphics[width=0.9\unitlength]{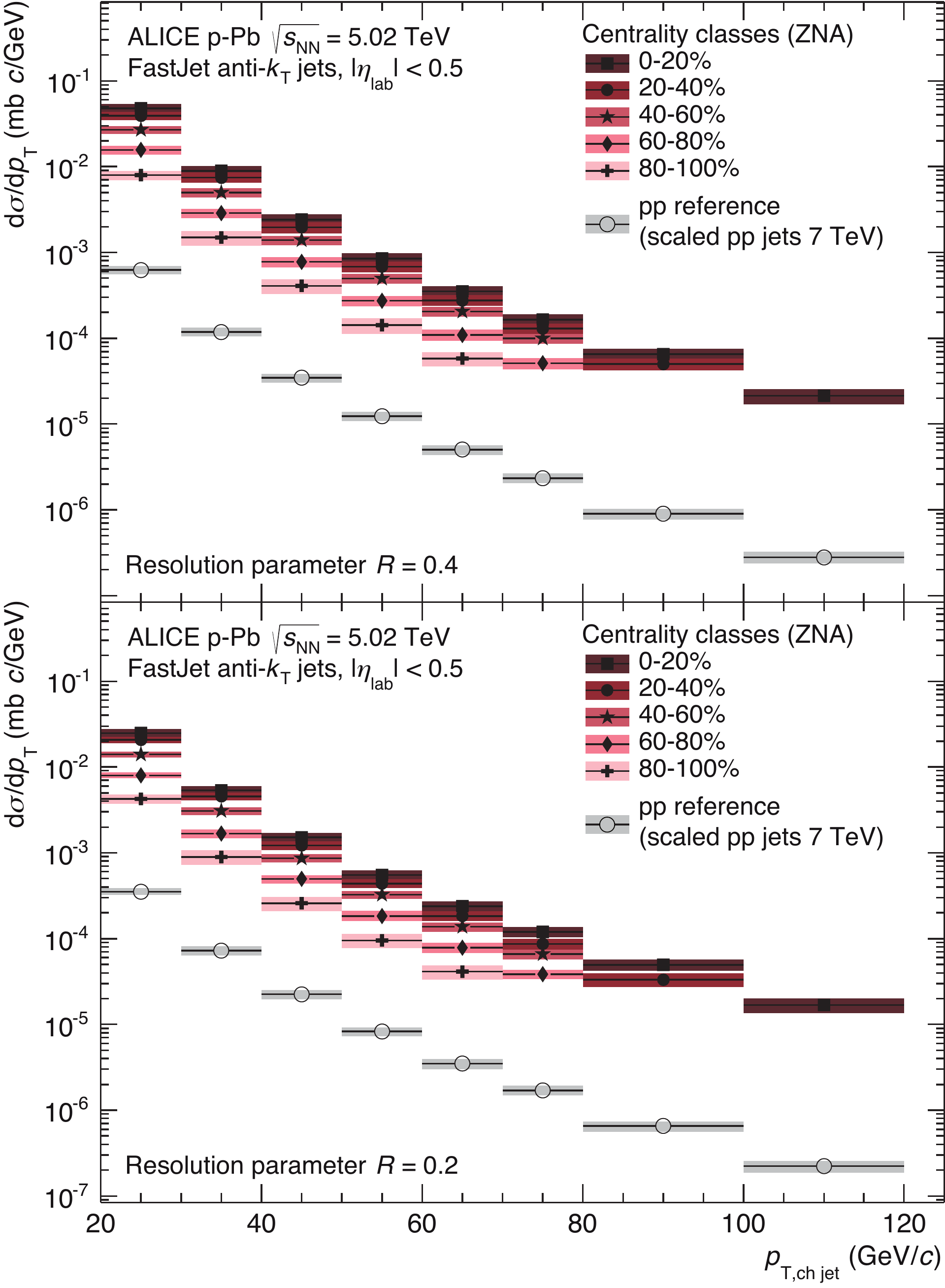}}
\end{picture}
\end{center}
\caption{(Color online) $\pt$-differential production cross sections
  of charged jet production in $\pPb$ collisions at $5.02$ TeV for
  several centrality classes. Top and bottom panels show the result
  for $R=0.4$ and $R=0.2$, respectively. In these and the following plots, the coloured boxes  
  represent systematic uncertainties, the error bars represent statistical uncertainties.
  The overall normalisation uncertainty on the visible cross section is $3.3\%$ in $\pPb$.
  The corresponding reference $\pp$ spectrum is shown for both radii, it was obtained 
  by scaling down the measured charged jets at 7 TeV to the reference energy.}
\label{fig:Spectra_ZNA}
\end{figure}

The $\pt$-differential cross sections for jets reconstructed from
charged particles  for five centrality classes in $\pPb$
collisions at $\snn = 5.02$ TeV are shown in Fig.~\ref{fig:Spectra_ZNA}.
For both resolution parameters, the measured yields are higher for more central collisions, as expected from the increase of the binary interactions (cf. Tab.~\ref{tab:cents_moments}). 
The $\pp$ reference at $\sqrt{s} = 5.02$~TeV is also shown. 
In addition to the increase in binary collisions the larger total cross section in $\pPb$ compared to $\pp$ further separates the data from the two collision systems; by an additional  factor of $20\% \cdot \sigma^{\mathrm{pPb}}_\mathrm{V0}/\sigma^\mathrm{pp}_\mathrm{inel} \approx  6$.

\begin{figure}[t]
\setlength{\unitlength}{0.75\columnwidth}
\begin{center}
\begin{picture}(1,1.25)
\put(0,0){\includegraphics[width=0.9\unitlength]{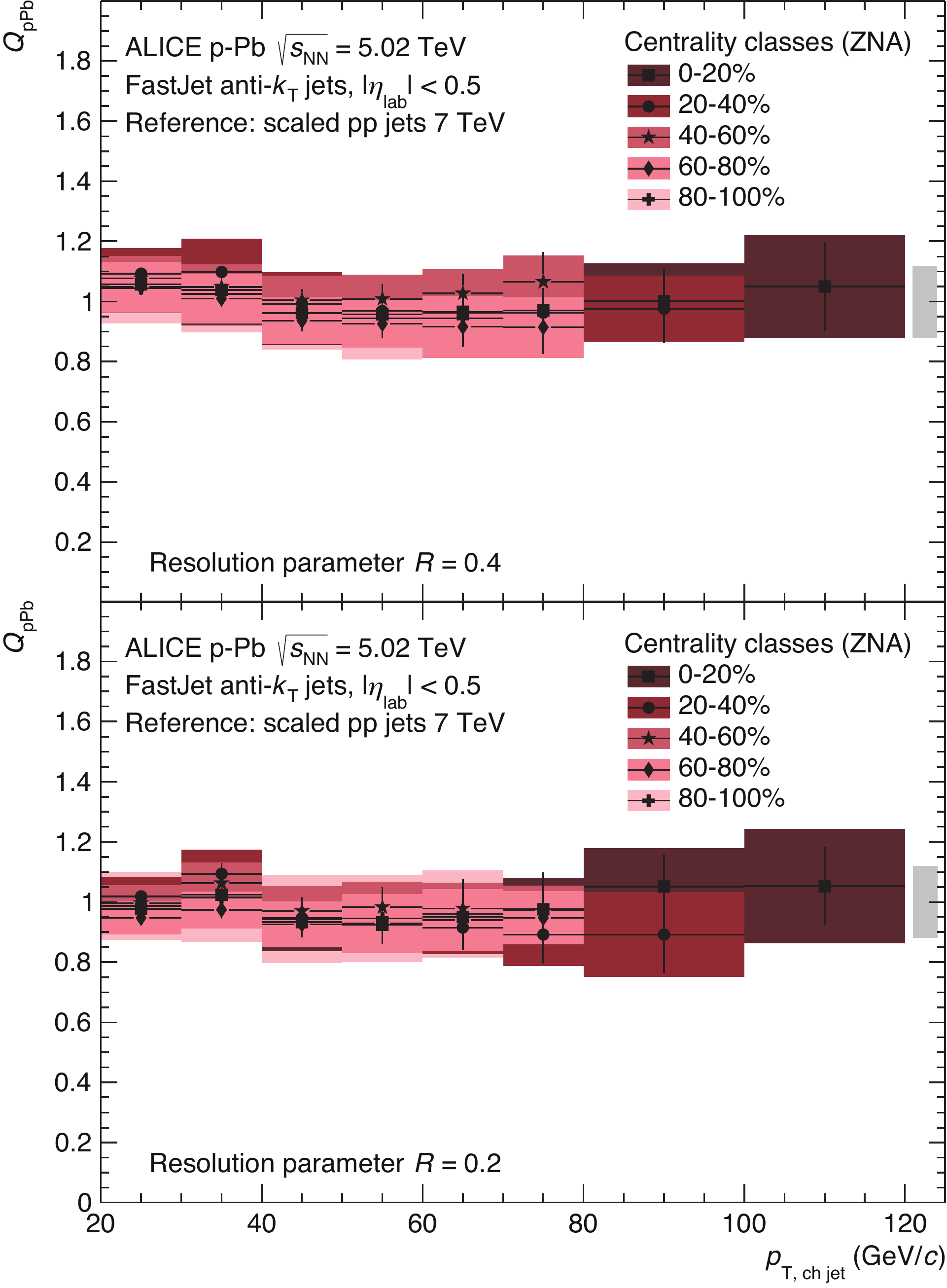}}
\end{picture}
\end{center}
\caption{(Color online) Nuclear modification factors $Q_\mathrm{pPb}$
  of charged jets for several centrality classes. $\ncoll$ has been determined with the hybrid model. Top and bottom
  panels show the result for $R=0.4$ and $R=0.2$, respectively. The
  combined global normalisation uncertainty from
  $\ncoll$, the measured pp cross section,
and the reference scaling is indicated by the box around unity.}
\label{fig:QpPb_ZNA_PbSide}
\end{figure}

The scaling behaviour of the $\pPb$ spectra with respect to the pp reference is quantified by the nuclear modification factor $\QpPb$ (Eq.\ref{eq:r_ppb}). 
The nuclear modification factor with the hybrid approach, shown in Fig.~\ref{fig:QpPb_ZNA_PbSide}, is compatible with unity for all centrality classes, indicating the absence of
centrality-dependent nuclear effects on the jet yield in the kinematic
regime probed by our measurement. 
This result is consistent with the measurement of single charged
particles in $\pPb$ collisions presented in \cite{Adam:2014qja}, where the same hybrid approach is used. 

For other centrality selections, closer to mid-rapidity, a separation of $\QpPb$ for jets is observed for the different centralities that is caused by dynamical biases of the selection, similar to the $\QpPb$ for charged particles. If we use e.g. the centrality selection based on the multiplicity in the V0A, $\QpPb$ decreases from about 1.2 in central to approximately 0.5 in peripheral collisions \cite{ALICE-PUBLIC-2016-XXX}. 

\begin{figure}
\begin{center}
\includegraphics[width=0.7\columnwidth]{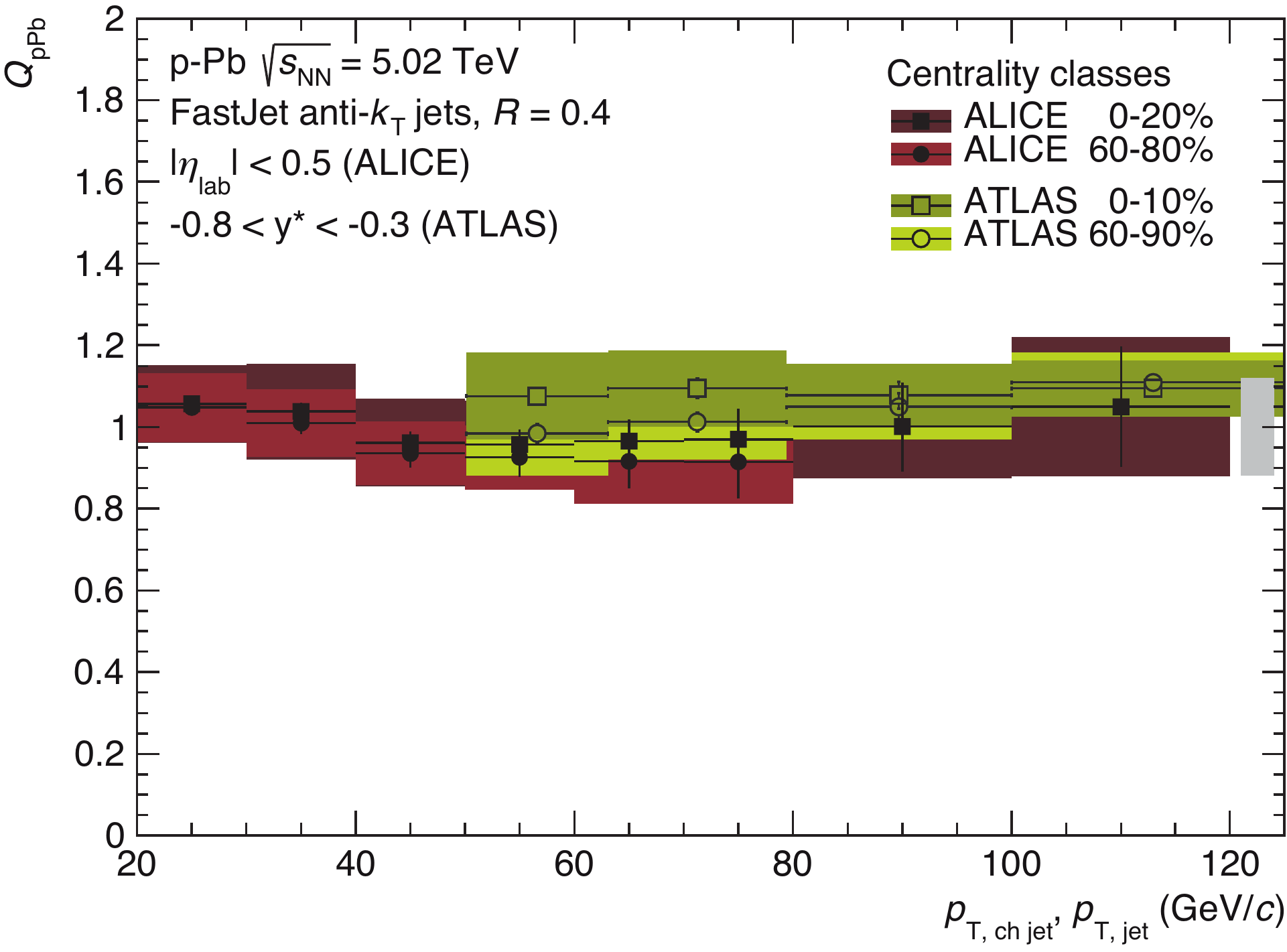}
\end{center}
\caption{(Color online) Nuclear modification factor of charged jets
  compared to the nuclear modification factor for full jets as
  measured by the ATLAS collaboration \cite{ATLAS:2014cpa}. Note that the underlying parton
  $\pt$ for fixed reconstructed jet $\pt$ is higher in the case of charged jets.}
\label{fig:compareATLAS}
\end{figure}

The centrality dependence of full jet production in $\pPb$ collisions, i.e.\ using charged and neutral jet fragments,
has been reported by the ATLAS collaboration in \cite{ATLAS:2014cpa} over a broad range of the
center-of-mass rapidity ($y^{*}$) and transverse momentum.
Centrality-dependent deviations of jet production have been found for large rapidities in the proton-going direction and $p_\mathrm{T, jet} \gtrsim 100$ GeV/$c$. 
In the nucleon--nucleon center-of-mass system as defined by ATLAS, our
measurement in $\left|\eta_\mathrm{lab}\right| < 0.5$ corresponds to $-0.96 < y^{*} < -0.04$.  
As shown in Fig.~\ref{fig:compareATLAS},  the measurement of the nuclear
modification factor of charged jets in central and peripheral
collisions is consistent with the full jet
measurement of ATLAS, where the kinematical selection of jet momentum
and rapidity overlap, note however that the underlying parton $\pT$ at a given reconstructed $\pt$ is higher for charged jets.

\begin{figure}
\begin{center}
\includegraphics[width=0.7\columnwidth]{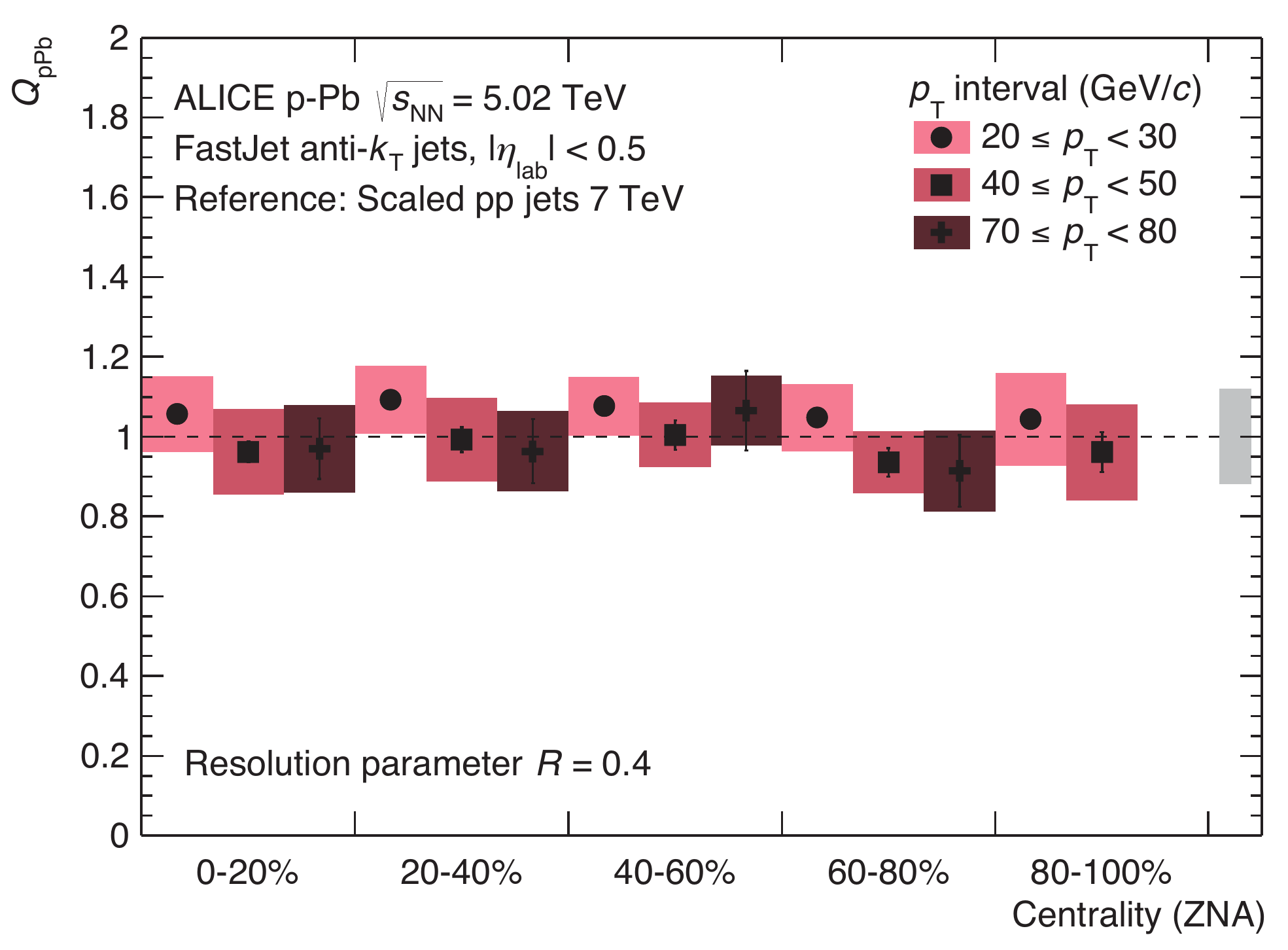}
\end{center}
\caption{(Color online) Centrality evolution of $\QpPb$ for selected
  $\ptch$-bins and $R = 0.4$. }
\label{fig:QpPb_ptvscent}
\end{figure}

The centrality evolution for $\QpPb$ as measured by ALICE is shown for three $\pt$-regions and $R = 0.4$ in Fig.~\ref{fig:QpPb_ptvscent}. 
No significant variation is observed with centrality for a fixed $\pt$ interval. The same holds for $R = 0.2$ (not shown).

Recently,   the PHENIX collaboration reported on a centrality
dependent modification of the jet yield in $\dAu$ collisions at $\snn
= 200$ GeV in the range of $20 <  \pt < 50~\mbox{GeV}/c$ \cite{Adare:2015gla}: 
a suppression of 20\% in central events and corresponding enhancement in
peripheral events is observed. 
Even when neglecting the impact of any possible biases in the
centrality selection, the measurement of the nuclear modification at lower $\snn$ cannot be directly compared to the measurements at LHC for two reasons. 
First, in case of a possible final state energy loss the scattered parton momentum is the
relevant scale. Here, the nuclear modification factor at
lower energies is more sensitive to energy loss, due to the steeper spectrum of
scattered partons.
Second, for initial state effects the nuclear modification should be compared
in the probed Bjorken-$x$,  which can be estimated at mid-rapidity to
$x_\mathrm{T} \approx 2\pt/\snn$, and is at a given $\pt$ 
approximately a factor of 25 smaller in $\pPb$ collisions at the LHC.

\begin{figure}
\begin{center}
\includegraphics[width=0.7\columnwidth]{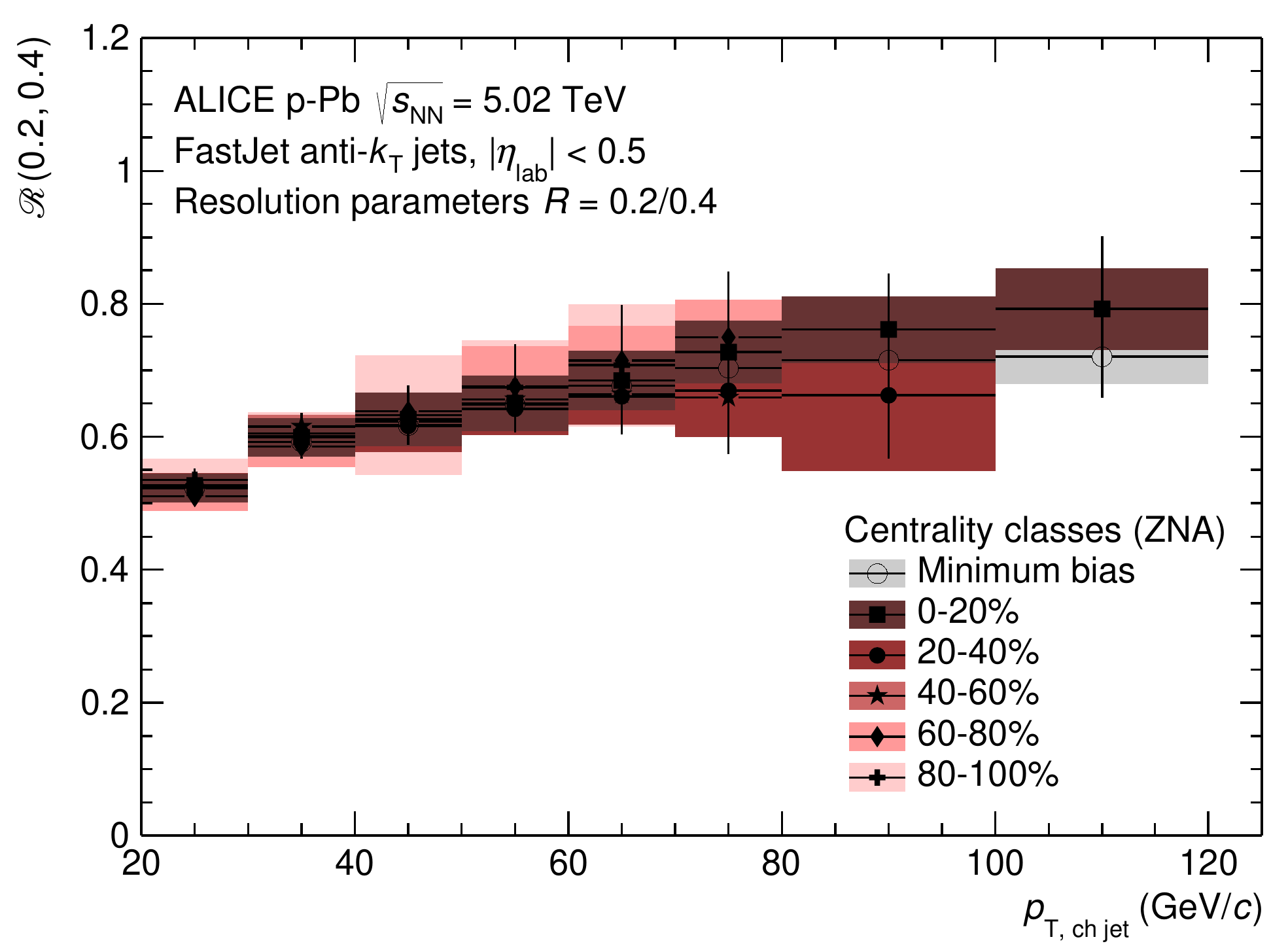}
\end{center}
\caption{(Color online) Charged jet production cross section ratio for
  different resolution parameters as defined in
  Eq.~\ref{eq:xsec_ratio}. Different centrality classes are
  shown together with the result for minimum bias collisions. Note that the systematic uncertainties are partially
  correlated between centrality classes. The ratio for minimum collisions is compared in more detail to  $\pp$ collisions at higher energy and NLO calculations at $\sqrt{s} = 5.02$ TeV  in \cite{Adam:2015hoa}, where no significant deviations are found.}
\label{fig:JetShape_ZNA}
\end{figure}

The ratio of jet production cross sections reconstructed
with $R = 0.2$ and 0.4 is shown in Fig.~\ref{fig:JetShape_ZNA}. 
For all centrality classes, the ratio shows the expected stronger jet collimation towards higher $\pt$. 
Moreover, the ratio is for all centralities consistent with the result obtained in minimum bias $\pPb$ collisions, which agrees with the jet cross section ratio in $\pp$ collisions as shown in \cite{Adam:2015hoa}.   
The result is fully compatible with the expectation, since even in
central $\PbPb$ collisions, where a significant jet
suppression in the nuclear modification factor is measured, the cross section
ratio remains unaffected \cite{Abelev:2013kqa}.
%

\section{Summary}
\label{sec:summary}

Centrality-dependent results on charged jet production in $\pPb$ collisions at $\sqrt{s_\mathrm{NN}} = 5.02$~TeV have been shown for transverse momentum range $20 <   p_{\mathrm{T,\,ch\;jet}} < 120~\mbox{GeV/}c$ and for resolution parameters $R=0.2$ and $R = 0.4$.
The centrality selection is performed using the forward neutron
energy, and the corresponding number of binary collisions $\ncoll$ is estimated via the correlation to the multiplicity measured in the lead-going direction, in order use a rapidity region  well separated from the one  where jets are reconstructed.

With this choice of centrality and data driven $\ncoll$ estimate, the nuclear modification factor $Q_\mathrm{pPb}$ is consistent with unity and does not indicate a
significant centrality dependence within the statistical and systematical uncertainties. 
In the measured kinematic range momentum between 20~$\mbox{GeV/}c$ and up to 120~$\mbox{GeV/}c$ and close to mid-rapidity, the observed nuclear modification factor  is consistent with
results from full jet measurements by the ATLAS collaboration in the same kinematic region.
The jet cross section ratio for  $R=0.2$ and $0.4$ shows no centrality dependence, indicating no modification of the degree of collimation of the jets at different centralities.

These measurements show the absence of strong nuclear effects on the jet production at mid-rapidity for all centralities.

\ifcernpp
\newenvironment{acknowledgement}{\relax}{\relax}
\begin{acknowledgement}
\section*{Acknowledgements}

The ALICE Collaboration would like to thank all its engineers and technicians for their invaluable contributions to the construction of the experiment and the CERN accelerator teams for the outstanding performance of the LHC complex.
The ALICE Collaboration gratefully acknowledges the resources and support provided by all Grid centres and the Worldwide LHC Computing Grid (WLCG) collaboration.
The ALICE Collaboration acknowledges the following funding agencies for their support in building and
running the ALICE detector:
State Committee of Science,  World Federation of Scientists (WFS)
and Swiss Fonds Kidagan, Armenia;
Conselho Nacional de Desenvolvimento Cient\'{\i}fico e Tecnol\'{o}gico (CNPq), Financiadora de Estudos e Projetos (FINEP),
Funda\c{c}\~{a}o de Amparo \`{a} Pesquisa do Estado de S\~{a}o Paulo (FAPESP);
National Natural Science Foundation of China (NSFC), the Chinese Ministry of Education (CMOE)
and the Ministry of Science and Technology of China (MSTC);
Ministry of Education and Youth of the Czech Republic;
Danish Natural Science Research Council, the Carlsberg Foundation and the Danish National Research Foundation;
The European Research Council under the European Community's Seventh Framework Programme;
Helsinki Institute of Physics and the Academy of Finland;
French CNRS-IN2P3, the `Region Pays de Loire', `Region Alsace', `Region Auvergne' and CEA, France;
German Bundesministerium fur Bildung, Wissenschaft, Forschung und Technologie (BMBF) and the Helmholtz Association;
General Secretariat for Research and Technology, Ministry of Development, Greece;
National Research, Development and Innovation Office (NKFIH), Hungary;
Department of Atomic Energy and Department of Science and Technology of the Government of India;
Istituto Nazionale di Fisica Nucleare (INFN) and Centro Fermi -
Museo Storico della Fisica e Centro Studi e Ricerche ``Enrico Fermi'', Italy;
Japan Society for the Promotion of Science (JSPS) KAKENHI and MEXT, Japan;
Joint Institute for Nuclear Research, Dubna;
National Research Foundation of Korea (NRF);
Consejo Nacional de Cienca y Tecnologia (CONACYT), Direccion General de Asuntos del Personal Academico(DGAPA), M\'{e}xico, Amerique Latine Formation academique - 
European Commission~(ALFA-EC) and the EPLANET Program~(European Particle Physics Latin American Network);
Stichting voor Fundamenteel Onderzoek der Materie (FOM) and the Nederlandse Organisatie voor Wetenschappelijk Onderzoek (NWO), Netherlands;
Research Council of Norway (NFR);
National Science Centre, Poland;
Ministry of National Education/Institute for Atomic Physics and National Council of Scientific Research in Higher Education~(CNCSI-UEFISCDI), Romania;
Ministry of Education and Science of Russian Federation, Russian
Academy of Sciences, Russian Federal Agency of Atomic Energy,
Russian Federal Agency for Science and Innovations and The Russian
Foundation for Basic Research;
Ministry of Education of Slovakia;
Department of Science and Technology, South Africa;
Centro de Investigaciones Energeticas, Medioambientales y Tecnologicas (CIEMAT), E-Infrastructure shared between Europe and Latin America (EELA), 
Ministerio de Econom\'{i}a y Competitividad (MINECO) of Spain, Xunta de Galicia (Conseller\'{\i}a de Educaci\'{o}n),
Centro de Aplicaciones Tecnológicas y Desarrollo Nuclear (CEA\-DEN), Cubaenerg\'{\i}a, Cuba, and IAEA (International Atomic Energy Agency);
Swedish Research Council (VR) and Knut $\&$ Alice Wallenberg
Foundation (KAW);
Ukraine Ministry of Education and Science;
United Kingdom Science and Technology Facilities Council (STFC);
The United States Department of Energy, the United States National
Science Foundation, the State of Texas, and the State of Ohio;
Ministry of Science, Education and Sports of Croatia and  Unity through Knowledge Fund, Croatia;
Council of Scientific and Industrial Research (CSIR), New Delhi, India;
Pontificia Universidad Cat\'{o}lica del Per\'{u}.
\end{acknowledgement}
\bibliographystyle{utphys} 
\else
\bibliographystyle{utphys} 
\fi
\bibliography{paper}
\newpage
%
%
\ifcernpp
\appendix
\section{The ALICE Collaboration}
\label{app:collab}



\begingroup
\small
\begin{flushleft}
J.~Adam\Irefn{org39}\And
D.~Adamov\'{a}\Irefn{org84}\And
M.M.~Aggarwal\Irefn{org88}\And
G.~Aglieri Rinella\Irefn{org35}\And
M.~Agnello\Irefn{org110}\And
N.~Agrawal\Irefn{org47}\And
Z.~Ahammed\Irefn{org133}\And
S.~Ahmad\Irefn{org19}\And
S.U.~Ahn\Irefn{org68}\And
S.~Aiola\Irefn{org137}
% ...
\renewcommand\labelenumi{\textsuperscript{\theenumi}~}

\section*{Affiliation notes}
\renewcommand\theenumi{\roman{enumi}}
\begin{Authlist}
\item \Adef{0}Deceased
\item \Adef{idp1778112}{Also at: Georgia State University, Atlanta, Georgia, United States}
\item \Adef{idp3133376}{Also at: Also at Department of Applied Physics, Aligarh Muslim University, Aligarh, India}
\item \Adef{idp3847728}{Also at: M.V. Lomonosov Moscow State University, D.V. Skobeltsyn Institute of Nuclear, Physics, Moscow, Russia}
\end{Authlist}

\section*{Collaboration Institutes}
\renewcommand\theenumi{\arabic{enumi}~}
\begin{Authlist}

\item \Idef{org1}A.I. Alikhanyan National Science Laboratory (Yerevan Physics Institute) Foundation, Yerevan, Armenia
\item \Idef{org2}Benem\'{e}rita Universidad Aut\'{o}noma de Puebla, Puebla, Mexico
\item \Idef{org3}Bogolyubov Institute for Theoretical Physics, Kiev, Ukraine
\item \Idef{org4}Bose Institute, Department of Physics and Centre for Astroparticle Physics and Space Science (CAPSS), Kolkata, India
\item \Idef{org5}Budker Institute for Nuclear Physics, Novosibirsk, Russia
\item \Idef{org6}California Polytechnic State University, San Luis Obispo, California, United States
\item \Idef{org7}Central China Normal University, Wuhan, China
\item \Idef{org8}Centre de Calcul de l'IN2P3, Villeurbanne, France
\item \Idef{org9}Centro de Aplicaciones Tecnol\'{o}gicas y Desarrollo Nuclear (CEADEN), Havana, Cuba
\item \Idef{org10}Centro de Investigaciones Energ\'{e}ticas Medioambientales y Tecnol\'{o}gicas (CIEMAT), Madrid, Spain
\item \Idef{org11}Centro de Investigaci\'{o}n y de Estudios Avanzados (CINVESTAV), Mexico City and M\'{e}rida, Mexico
\item \Idef{org12}Centro Fermi - Museo Storico della Fisica e Centro Studi e Ricerche ``Enrico Fermi'', Rome, Italy
\item \Idef{org13}Chicago State University, Chicago, Illinois, USA
\item \Idef{org14}China Institute of Atomic Energy, Beijing, China
\item \Idef{org15}Commissariat \`{a} l'Energie Atomique, IRFU, Saclay, France
\item \Idef{org16}COMSATS Institute of Information Technology (CIIT), Islamabad, Pakistan
\item \Idef{org17}Departamento de F\'{\i}sica de Part\'{\i}culas and IGFAE, Universidad de Santiago de Compostela, Santiago de Compostela, Spain
\item \Idef{org18}Department of Physics and Technology, University of Bergen, Bergen, Norway
\item \Idef{org19}Department of Physics, Aligarh Muslim University, Aligarh, India
\item \Idef{org20}Department of Physics, Ohio State University, Columbus, Ohio, United States
\item \Idef{org21}Department of Physics, Sejong University, Seoul, South Korea
\item \Idef{org22}Department of Physics, University of Oslo, Oslo, Norway
\item \Idef{org23}Dipartimento di Fisica dell'Universit\`{a} 'La Sapienza' and Sezione INFN Rome, Italy
\item \Idef{org24}Dipartimento di Fisica dell'Universit\`{a} and Sezione INFN, Cagliari, Italy
\item \Idef{org25}Dipartimento di Fisica dell'Universit\`{a} and Sezione INFN, Trieste, Italy
\item \Idef{org26}Dipartimento di Fisica dell'Universit\`{a} and Sezione INFN, Turin, Italy
\item \Idef{org27}Dipartimento di Fisica e Astronomia dell'Universit\`{a} and Sezione INFN, Bologna, Italy
\item \Idef{org28}Dipartimento di Fisica e Astronomia dell'Universit\`{a} and Sezione INFN, Catania, Italy
\item \Idef{org29}Dipartimento di Fisica e Astronomia dell'Universit\`{a} and Sezione INFN, Padova, Italy
\item \Idef{org30}Dipartimento di Fisica `E.R.~Caianiello' dell'Universit\`{a} and Gruppo Collegato INFN, Salerno, Italy
\item \Idef{org31}Dipartimento di Scienze e Innovazione Tecnologica dell'Universit\`{a} del  Piemonte Orientale and Gruppo Collegato INFN, Alessandria, Italy
\item \Idef{org32}Dipartimento Interateneo di Fisica `M.~Merlin' and Sezione INFN, Bari, Italy
\item \Idef{org33}Division of Experimental High Energy Physics, University of Lund, Lund, Sweden
\item \Idef{org34}Eberhard Karls Universit\"{a}t T\"{u}bingen, T\"{u}bingen, Germany
\item \Idef{org35}European Organization for Nuclear Research (CERN), Geneva, Switzerland
\item \Idef{org36}Excellence Cluster Universe, Technische Universit\"{a}t M\"{u}nchen, Munich, Germany
\item \Idef{org37}Faculty of Engineering, Bergen University College, Bergen, Norway
\item \Idef{org38}Faculty of Mathematics, Physics and Informatics, Comenius University, Bratislava, Slovakia
\item \Idef{org39}Faculty of Nuclear Sciences and Physical Engineering, Czech Technical University in Prague, Prague, Czech Republic
\item \Idef{org40}Faculty of Science, P.J.~\v{S}af\'{a}rik University, Ko\v{s}ice, Slovakia
\item \Idef{org41}Faculty of Technology, Buskerud and Vestfold University College, Vestfold, Norway
\item \Idef{org42}Frankfurt Institute for Advanced Studies, Johann Wolfgang Goethe-Universit\"{a}t Frankfurt, Frankfurt, Germany
\item \Idef{org43}Gangneung-Wonju National University, Gangneung, South Korea
\item \Idef{org44}Gauhati University, Department of Physics, Guwahati, India
\item \Idef{org45}Helsinki Institute of Physics (HIP), Helsinki, Finland
\item \Idef{org46}Hiroshima University, Hiroshima, Japan
\item \Idef{org47}Indian Institute of Technology Bombay (IIT), Mumbai, India
\item \Idef{org48}Indian Institute of Technology Indore, Indore (IITI), India
\item \Idef{org49}Indonesian Institute of Sciences, Jakarta, Indonesia
\item \Idef{org50}Inha University, Incheon, South Korea
\item \Idef{org51}Institut de Physique Nucl\'eaire d'Orsay (IPNO), Universit\'e Paris-Sud, CNRS-IN2P3, Orsay, France
\item \Idef{org52}Institut f\"{u}r Informatik, Johann Wolfgang Goethe-Universit\"{a}t Frankfurt, Frankfurt, Germany
\item \Idef{org53}Institut f\"{u}r Kernphysik, Johann Wolfgang Goethe-Universit\"{a}t Frankfurt, Frankfurt, Germany
\item \Idef{org54}Institut f\"{u}r Kernphysik, Westf\"{a}lische Wilhelms-Universit\"{a}t M\"{u}nster, M\"{u}nster, Germany
\item \Idef{org55}Institut Pluridisciplinaire Hubert Curien (IPHC), Universit\'{e} de Strasbourg, CNRS-IN2P3, Strasbourg, France
\item \Idef{org56}Institute for Nuclear Research, Academy of Sciences, Moscow, Russia
\item \Idef{org57}Institute for Subatomic Physics of Utrecht University, Utrecht, Netherlands
\item \Idef{org58}Institute for Theoretical and Experimental Physics, Moscow, Russia
\item \Idef{org59}Institute of Experimental Physics, Slovak Academy of Sciences, Ko\v{s}ice, Slovakia
\item \Idef{org60}Institute of Physics, Academy of Sciences of the Czech Republic, Prague, Czech Republic
\item \Idef{org61}Institute of Physics, Bhubaneswar, India
\item \Idef{org62}Institute of Space Science (ISS), Bucharest, Romania
\item \Idef{org63}Instituto de Ciencias Nucleares, Universidad Nacional Aut\'{o}noma de M\'{e}xico, Mexico City, Mexico
\item \Idef{org64}Instituto de F\'{\i}sica, Universidad Nacional Aut\'{o}noma de M\'{e}xico, Mexico City, Mexico
\item \Idef{org65}iThemba LABS, National Research Foundation, Somerset West, South Africa
\item \Idef{org66}Joint Institute for Nuclear Research (JINR), Dubna, Russia
\item \Idef{org67}Konkuk University, Seoul, South Korea
\item \Idef{org68}Korea Institute of Science and Technology Information, Daejeon, South Korea
\item \Idef{org69}KTO Karatay University, Konya, Turkey
\item \Idef{org70}Laboratoire de Physique Corpusculaire (LPC), Clermont Universit\'{e}, Universit\'{e} Blaise Pascal, CNRS--IN2P3, Clermont-Ferrand, France
\item \Idef{org71}Laboratoire de Physique Subatomique et de Cosmologie, Universit\'{e} Grenoble-Alpes, CNRS-IN2P3, Grenoble, France
\item \Idef{org72}Laboratori Nazionali di Frascati, INFN, Frascati, Italy
\item \Idef{org73}Laboratori Nazionali di Legnaro, INFN, Legnaro, Italy
\item \Idef{org74}Lawrence Berkeley National Laboratory, Berkeley, California, United States
\item \Idef{org75}Moscow Engineering Physics Institute, Moscow, Russia
\item \Idef{org76}Nagasaki Institute of Applied Science, Nagasaki, Japan
\item \Idef{org77}National Centre for Nuclear Studies, Warsaw, Poland
\item \Idef{org78}National Institute for Physics and Nuclear Engineering, Bucharest, Romania
\item \Idef{org79}National Institute of Science Education and Research, Bhubaneswar, India
\item \Idef{org80}National Research Centre Kurchatov Institute, Moscow, Russia
\item \Idef{org81}Niels Bohr Institute, University of Copenhagen, Copenhagen, Denmark
\item \Idef{org82}Nikhef, Nationaal instituut voor subatomaire fysica, Amsterdam, Netherlands
\item \Idef{org83}Nuclear Physics Group, STFC Daresbury Laboratory, Daresbury, United Kingdom
\item \Idef{org84}Nuclear Physics Institute, Academy of Sciences of the Czech Republic, \v{R}e\v{z} u Prahy, Czech Republic
\item \Idef{org85}Oak Ridge National Laboratory, Oak Ridge, Tennessee, United States
\item \Idef{org86}Petersburg Nuclear Physics Institute, Gatchina, Russia
\item \Idef{org87}Physics Department, Creighton University, Omaha, Nebraska, United States
\item \Idef{org88}Physics Department, Panjab University, Chandigarh, India
\item \Idef{org89}Physics Department, University of Athens, Athens, Greece
\item \Idef{org90}Physics Department, University of Cape Town, Cape Town, South Africa
\item \Idef{org91}Physics Department, University of Jammu, Jammu, India
\item \Idef{org92}Physics Department, University of Rajasthan, Jaipur, India
\item \Idef{org93}Physik Department, Technische Universit\"{a}t M\"{u}nchen, Munich, Germany
\item \Idef{org94}Physikalisches Institut, Ruprecht-Karls-Universit\"{a}t Heidelberg, Heidelberg, Germany
\item \Idef{org95}Purdue University, West Lafayette, Indiana, United States
\item \Idef{org96}Pusan National University, Pusan, South Korea
\item \Idef{org97}Research Division and ExtreMe Matter Institute EMMI, GSI Helmholtzzentrum f\"ur Schwerionenforschung, Darmstadt, Germany
\item \Idef{org98}Rudjer Bo\v{s}kovi\'{c} Institute, Zagreb, Croatia
\item \Idef{org99}Russian Federal Nuclear Center (VNIIEF), Sarov, Russia
\item \Idef{org100}Saha Institute of Nuclear Physics, Kolkata, India
\item \Idef{org101}School of Physics and Astronomy, University of Birmingham, Birmingham, United Kingdom
\item \Idef{org102}Secci\'{o}n F\'{\i}sica, Departamento de Ciencias, Pontificia Universidad Cat\'{o}lica del Per\'{u}, Lima, Peru
\item \Idef{org103}Sezione INFN, Bari, Italy
\item \Idef{org104}Sezione INFN, Bologna, Italy
\item \Idef{org105}Sezione INFN, Cagliari, Italy
\item \Idef{org106}Sezione INFN, Catania, Italy
\item \Idef{org107}Sezione INFN, Padova, Italy
\item \Idef{org108}Sezione INFN, Rome, Italy
\item \Idef{org109}Sezione INFN, Trieste, Italy
\item \Idef{org110}Sezione INFN, Turin, Italy
\item \Idef{org111}SSC IHEP of NRC Kurchatov institute, Protvino, Russia
\item \Idef{org112}Stefan Meyer Institut f\"{u}r Subatomare Physik (SMI), Vienna, Austria
\item \Idef{org113}SUBATECH, Ecole des Mines de Nantes, Universit\'{e} de Nantes, CNRS-IN2P3, Nantes, France
\item \Idef{org114}Suranaree University of Technology, Nakhon Ratchasima, Thailand
\item \Idef{org115}Technical University of Ko\v{s}ice, Ko\v{s}ice, Slovakia
\item \Idef{org116}Technical University of Split FESB, Split, Croatia
\item \Idef{org117}The Henryk Niewodniczanski Institute of Nuclear Physics, Polish Academy of Sciences, Cracow, Poland
\item \Idef{org118}The University of Texas at Austin, Physics Department, Austin, Texas, USA
\item \Idef{org119}Universidad Aut\'{o}noma de Sinaloa, Culiac\'{a}n, Mexico
\item \Idef{org120}Universidade de S\~{a}o Paulo (USP), S\~{a}o Paulo, Brazil
\item \Idef{org121}Universidade Estadual de Campinas (UNICAMP), Campinas, Brazil
\item \Idef{org122}University of Houston, Houston, Texas, United States
\item \Idef{org123}University of Jyv\"{a}skyl\"{a}, Jyv\"{a}skyl\"{a}, Finland
\item \Idef{org124}University of Liverpool, Liverpool, United Kingdom
\item \Idef{org125}University of Tennessee, Knoxville, Tennessee, United States
\item \Idef{org126}University of the Witwatersrand, Johannesburg, South Africa
\item \Idef{org127}University of Tokyo, Tokyo, Japan
\item \Idef{org128}University of Tsukuba, Tsukuba, Japan
\item \Idef{org129}University of Zagreb, Zagreb, Croatia
\item \Idef{org130}Universit\'{e} de Lyon, Universit\'{e} Lyon 1, CNRS/IN2P3, IPN-Lyon, Villeurbanne, France
\item \Idef{org131}Universit\`{a} di Brescia
\item \Idef{org132}V.~Fock Institute for Physics, St. Petersburg State University, St. Petersburg, Russia
\item \Idef{org133}Variable Energy Cyclotron Centre, Kolkata, India
\item \Idef{org134}Warsaw University of Technology, Warsaw, Poland
\item \Idef{org135}Wayne State University, Detroit, Michigan, United States
\item \Idef{org136}Wigner Research Centre for Physics, Hungarian Academy of Sciences, Budapest, Hungary
\item \Idef{org137}Yale University, New Haven, Connecticut, United States
\item \Idef{org138}Yonsei University, Seoul, South Korea
\item \Idef{org139}Zentrum f\"{u}r Technologietransfer und Telekommunikation (ZTT), Fachhochschule Worms, Worms, Germany
\end{Authlist}
\endgroup

\fi

\end{document}